\documentclass[3p,twocolumn,number,times]{elsarticle}

\usepackage{graphicx}
\usepackage{amsmath}
\usepackage{amsthm}
\usepackage{amssymb}
\usepackage{amsfonts}

\begin{document}

\title{Boundary effects on quantum q-breathers in a Bose-Hubbard chain}

\author[mpi]{Ricardo A. Pinto\corref{cor1}}
\ead{pinto@mpipks-dresden.mpg.de}
\author[mpi,fpl]{Jean Pierre Nguenang}
\ead{nguenang@yahoo.com}
\author[mpi]{Sergej Flach}
\ead{flach@mpipks-dresden.mpg.de}

\cortext[cor1]{Corresponding author}
\address[mpi]{Max-Planck-Institut f\"ur Physik komplexer Systeme, N\"othnitzer Str. 38, 01187 Dresden, Germany}
\address[fpl]{Fundamental physics laboratory: Group of Nonlinear physics and Complex
systems, Department of Physics, University of Douala,
 P.O. Box 24157, Douala-Cameroon}

\date{\today}

\begin{abstract}

We investigate the spectrum and eigenstates of a
Bose-Hubbard chain containing two bosons with fixed boundary conditions.
In the noninteracting case the eigenstates of the system define a two-dimensional normal-mode space.
For the interacting case weight functions of the eigenstates are computed by perturbation theory and numerical diagonalization. 
We identify paths in the two-dimensional normal-mode space which are rims for the weight functions.
The decay along and off the rims is algebraic.
Intersection of two 
paths (rims) leads to a local enhancement of the
weight functions. We analyze nonperturbative effects due to the degeneracies and
the formation of two-boson bound states.

\end{abstract}


\maketitle

\section{Introduction}

Localization phenomena due to nonlinearity and spatial discreteness in different 
physical systems have received considerable interest during the past few decades.
Despite the given translational invariance of a lattice, nonlinearity may trap
initially localized excitations.
The generic existence and properties of discrete breathers - time-periodic and spatially localized solutions 
of the underlying classical equations of motion - allow us to describe and understand
these localization phenomena  
\cite{FlachPhysRep295,physicstoday,Sievers,AubryPhysicaD103}.
Discrete breathers were observed in many different systems
like bond 
excitations in molecules, lattice vibrations and spin excitations in solids,
electronic currents in coupled Josephson junctions, light propagation in
interacting optical waveguides, cantilever vibrations in micromechanical
arrays, cold atom dynamics in Bose-Einstein condensates loaded on optical
lattices, among others (for references see \cite{FlachPhysRep295,physicstoday}). 
In many cases quantum effects are important.
Quantum breathers are nearly degenerate many-quanta
bound states which, when superposed, form a spatially localized excitation with a very long time to 
tunnel from one lattice site to another (for references see
\cite{FlachPhysRep295,physicstoday,AubryPhysicaD103}).

The application of the above ideas to normal-mode space of a classical nonlinear lattice allowed us
to explain many facets of the Fermi-Pasta-Ulam (FPU) paradox \cite{Fermi},
which consists of the nonequipartition of energy among the linear normal
modes in a nonlinear chain. There, the energy stays trapped in the initially excited
normal mode with only a few other normal modes excited, leading to localization of energy
in normal-mode space. Recent studies showed that, similar to discrete breathers,
exact time-periodic orbits exist which are localized in normal-mode space.
The properties of these $q$-breathers 
\cite{FlachPRL95,IvanchenkoPRL2006,FlachPRE2006,oik07,tp07,sfap08,kgm08,sf08} 
allow us to quantitatively address the observations of 
the FPU paradox. A hallmark of $q$-breathers is the exponential localization of energy 
in normal-mode space, with exponents depending on control parameters of the system.

On the quantum side, recently we studied the fate of analogous states 
(quantum $q$-breathers) in a one-dimensional lattice
with two interacting bosons and periodic boundary conditions \cite{NguenangPRB75}. 
By using perturbation theory, 
supported by numerical diagonalization, we computed weight functions of the eigenstates of the system 
in the many-body normal-mode space. 
We did find localization of the weight function in normal-mode space. However, 
at variance from the classical case, the decay is algebraic instead of exponential.
The periodic boundary conditions allow us to introduce an irreducible Bloch representation. 
Since states with different wave numbers belong to different Hilbert subspaces, 
they are not coupled by a Hubbard interaction term. Therefore, localization along the
Bloch wave number is compact. This is also happening for the corresponding classical
nonlinear Schr\"odinger equation with periodic boundary conditions \cite{kgm08},
when searching for plane-wave-like states.

The classical case however inevitably
leads to noncompact distributions in normal-mode space, once fixed boundary conditions are considered.
Indeed, also in the quantum case, these conditions violate translational invariance,
and lead to nonzero matrix elements between states with different Bloch wave numbers, mediated
by the Hubbard interaction. That is the reason for studying the properties of
quantum $q$-breathers for finite chains with fixed boundary conditions.
From a technical point of view, the irreducible normal-mode space dimension is then
increased from one to two.

In Sec. \ref{model} we describe the model and introduce the basis to write 
down the Hamiltonian matrix. We describe the quantum states of the lattice containing 
one and two noninteracting bosons. From the latter case we use the two-particle states as the basis to 
write down the Hamiltonian matrix in normal-mode space for the interacting case, after which the 
energy spectrum is computed. In Sec. \ref{localization} we study localization in normal-mode space. 
We introduce weight functions to describe localization in that space, and obtain analytical 
predictions using
perturbation theory. We present numerical results from a diagonalization of the Hamiltonian matrix, and compare them with analytical
estimates. Then we study nonperturbative effects when increasing the interaction parameter. Finally we present our conclusions in Sec. \ref{conclusions}.

\section{Model and spectrum}\label{model}

We consider a one-dimensional periodic lattice with
$f$ sites described by the Bose-Hubbard (BH) model. This is a quantum version
of the discrete nonlinear Schr\"odinger equation, which has been used to
describe a great variety of systems \cite{Scott1}. The BH Hamiltonian is $\hat{H} = \hat{H}_0 + \gamma \hat{H}_1$ \cite{EilbeckPhysicaD78}, with
\begin{equation}\label{eq:hamiltonian}
\hat{H}_0=-\sum_{j=1}^{f} \hat{a}_j^+(\hat{a}_{j-1}+\hat{a}_{j+1}),
\end{equation}
and
\begin{equation}\label{eq:haminteraction}
\hat{H}_1 = - \sum_{j=1}^f \hat{a}_j^+\hat{a}_j^+\hat{a}_j\hat{a}_j.
\end{equation}
$\hat{H}_0$ describes the nearest-neighbor hopping of particles (bosons) along the lattice, 
and $\hat{H}_1$ the local interaction between them whose strength is controlled by the parameter $\gamma$.
$ a_j^+$ and $a_j$ are the bosonic creation and annihilation operators satisfying the commutation 
relations $[\hat{a}_j,\hat{a}_{j'}^+]=\delta_{j,j'}$, $[\hat{a}_j,\hat{a}_{j'}]=[\hat{a}_j^+,\hat{a}_{j'}^+]=0$, 
and the system is subject to fixed boundary conditions. The Hamiltonian (\ref{eq:hamiltonian})
commutes with the number operator $\hat{N}=\sum_{j=1}^{f}{\hat{a}_j^+\hat{a}_j}$ whose
eigenvalue is $n$, the total number of bosons in the lattice. Here $n=2$.
It is of interest due to its direct relevance to studies and observation of two-vibron 
bound states in molecules and solids 
\cite{Cohen1969,Kimball,Richter1988,GuyotSionnest1991,Dai1994,Chin1995,Jakob1996,JakobPr75,
Pouthier2003JCP,Okuyama2001,Pouthier2003PRE,Edler2004,Proville2005EurophysLett69,Proville2005PRB71,
Ivic2006PhysicaD216}. More recently, two-boson bound states have been observed in Bose-Einstein 
condensates loaded on an optical lattice \cite{Winkler2006Nature441}.

To describe quantum states, we use a number state basis $|\Phi_n\rangle=|n_1\; n_2\cdots n_f\rangle$ 
\cite{EilbeckPhysicaD78}, where $n_i=0,1,2$ represents the number of bosons at the i-th site of the lattice.
$|\Phi_n\rangle$ is an eigenstate of the number operator $\hat{N}$ with eigenvalue $n=\sum_{j=1}^f n_j$.

\subsection{One-particle states}

For the case of having only one boson in the lattice ($n=1$) a number state has the form 
$|0\cdots 0\;1_l\;0\;0\cdots 0\rangle \equiv |l\rangle$, where $l$ denotes the lattice site where 
the boson is. This number state can be also written as
\begin{equation}\label{eq:lstate}
|l\rangle=\hat{a}_l^+|0\rangle,
\end{equation}
where the operator $\hat{a}_l^+$ creates a boson at the $l$-th site of the lattice, and $|0\rangle$ is 
the vacuum state. 

We write down the Hamiltonian matrix in the basis of the above-defined number states. For the 
single-boson case, the interaction term $\hat{H}_1$ has no contribution to the matrix elements. 
The eigenstates of $\hat{H}_0$, for fixed boundary conditions, are standing waves:
\begin{eqnarray}\label{eq:kstate}
|\Psi_k\rangle = \sum_{l=1}^f  \sqrt{\frac{2}{f+1}}\sin\left(kl\right) |l\rangle 
\equiv  |k\rangle,
\end{eqnarray}
where $k = q\pi/(f+1)$, and $q=1,\ldots,f$.
The corresponding eigenenergies are
\begin{equation}\label{eq:1bosonenergy}
\varepsilon_k=-2\cos(k).
\end{equation}

We define bosonic operators $\hat{a}_k$, $\hat{a}_k^+$ satisfying the commutation relations 
$[\hat{a}_k,\hat{a}_{k'}^+]=\delta_{k,k'}$, $[\hat{a}_k,\hat{a}_{k'}]=[\hat{a}_k^+,\hat{a}_{k'}^+]=0$, 
such that the state (\ref{eq:kstate}) may be written similar to (\ref{eq:lstate}):
\begin{equation}\label{eq:kstate2}
|k\rangle=\hat{a}_k^+|0\rangle\;,\;
\hat{a}_{k}^+ = \sum_{l=1}^fS_{l,k}\hat{a}_{l}^+,
\end{equation}
where the operator $\hat{a}_k^+$ creates a boson in the single-particle state with quantum number 
(wave number or momentum) $k$. 
The bosonic operators $\hat{a}_k$, $\hat{a}_k^+$ are related to the operators $\hat{a}_l$, $\hat{a}_l^+$ 
in direct space through the transformation matrix
\begin{equation}\label{eq:Slk}
S_{l,k}=\sqrt{\frac{2}{f+1}}\sin(kl).
\end{equation}

\subsection{Two-particle states}\label{twobosonstates}

For the two-boson case ($n=2$), we define the number state basis in a similar way as in the 
single-boson case:
\begin{equation}\label{eq:l1l2}
|l_1,l_2\rangle= \sqrt{\frac{2 - \delta_{l_1,l_2}}{2}} \; \hat{a}_{l_1}^+\hat{a}_{l_2}^+|0\rangle,
\end{equation}
where $l_2\geq l_1$ because of the indistinguishability of particles. $\hat{a}_{l_1}^+$ and $\hat{a}_{l_2}^+$ 
respectively create one boson at the lattice sites $l_1$ and $l_2$. The number 
of basis states is $d= f(f+1)/2$. The interaction term 
$\hat{H}_1$ in (\ref{eq:hamiltonian}) contributes to the matrix elements of the 
Hamiltonian in the above-defined basis.

In the noninteracting case ($\gamma=0$) the eigenstates of $\hat{H}$ in terms of bosonic 
operators in the normal-mode space read [see Eq. (\ref{eq:kstate2})]:
\begin{equation}\label{eq:k1k2}
|k_1,k_2\rangle = \sqrt{\frac{2 - \delta_{q_1,q_2}}{2}} \; \hat{a}_{k_1}^+\hat{a}_{k_2}^+|0\rangle\;,\;
q_2\geq q_1.
\end{equation}
$\hat{a}_{k_1}^+$ and $\hat{a}_{k_2}^+$ respectively create one boson in the single-particle states 
$k_1$ and $k_2$ of the form (\ref{eq:kstate}).
Using Eqs. (\ref{eq:kstate2}) and (\ref{eq:Slk}), the relation between the basis states in 
normal-mode space (\ref{eq:k1k2}) and the basis states in direct space (\ref{eq:l1l2}) reads:
\begin{eqnarray}\label{eq:basis1}
|k_1,k_2\rangle &=& \frac{\sqrt{2-\delta_{{q}_1,{q}_2}}}{\sqrt{2}} \nonumber \\
                &\times & \Bigg[{\sum_{{l}_1=1}^f}{\sum_{{l}_2>{l}_1}^f}(S_{{l}_1,{k}_1}S_{l_2,k_2}
+ S_{l_2,k_1}S_{l_1,k_2})|l_1,l_2\rangle \nonumber \\
& & + \sqrt{2}\;{\sum_{{l}=1}^f}S_{l,{k}_1}S_{l,{k}_2}|l,l\rangle \Bigg].
\end{eqnarray}
In the interacting case ($\gamma>0$), we represent the eigenstates of the Hamiltonian (\ref{eq:hamiltonian}) in the 
normal-mode basis (\ref{eq:basis1}) of the noninteracting case. This leads to a $d\times d$ matrix 
[$d=f(f+1)/2$] whose elements $H(i,j)$ ($i,j=1,\ldots,d$) are
\begin{equation}\label{eq:matel}
H(i,j)=\langle k'_1,k'_2|\hat{H}|k_1,k_2\rangle \equiv \langle q'_1,q'_2|\hat{H}|q_1,q_2\rangle.
\end{equation}  
The integer $j$ that labels the column of the matrix element (\ref{eq:matel}) is related to the mode numbers $q_1$ and $q_2$ by
\begin{equation}\label{eq:indexj} 
j_{q_1,q_2}=(q_1 - 1)(f+1) -\frac{(q_1-1)(q_1+2)}{2}+ q_2.
\end{equation}
The same relation holds for the integer $i_{q'_1,q'_2}$ labeling the row of the matrix element (\ref{eq:matel}).

The matrix elements (\ref{eq:matel}) are
\begin{equation}\label{eq:matel2}
H(i,j) = H_0(i,j) + \gamma H_1(i,j),
\end{equation}
where
\begin{equation}\label{eq:matelem2}
H_0(i,j)= ( \varepsilon_{k_1} + \varepsilon_{k_2} )\delta_{i,j},
\end{equation}
and
\begin{equation}\label{eq:matelem1}
H_1(i,j) =f_{q_1,q_2,q'_1,q'_2}\sum_{l=1}^fS_{l,k_1}S_{l,k_2}S_{l,k'_1}S_{l,k'_2}.
\end{equation}
$\varepsilon_k$ is the single-particle energy given by Eq. (\ref{eq:1bosonenergy}), and the coefficients $f_{q_1,q_2,q'_1,q'_2}$ are
\begin{equation}\label{eq:fq}
f_{q_1,q_2,q'_1,q'_2}=-\frac{8\sqrt{(2-{\delta}_{q_1,q_2})(2-{\delta}_{q'_1,q'_2})}}{(f+1)^2}.
\end{equation} 
\begin{figure}[!b]
\begin{center}
\includegraphics[width=3.1in]{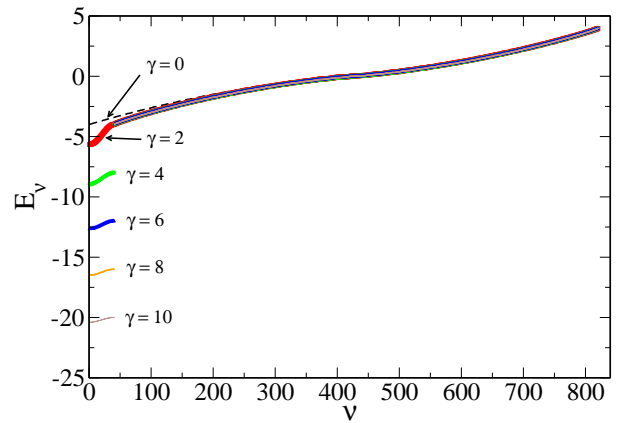}
\end{center}
\caption{\label{spectrum}Energy spectrum of the two-boson BH chain with fixed
boundary conditions for different values of the interaction strength $\gamma$. The eigenvalues are plotted as a function of the eigenvalue label (see text). Here $f=40$.}
\end{figure} 

In Fig. \ref{spectrum} we show the energy spectrum of the Hamiltonian matrix (\ref{eq:matel2}) 
obtained by numerical diagonalization for different values of the interaction parameter $\gamma$. 
In all calculations by numerical diagonalization we used $f=40$, which leads to a matrix dimension $d=820$. 
The eigenstates are ordered with respect to their eigenvalues $E_{\nu}$ ($\nu=1,\ldots,d$).
At $\gamma=0$, the spectrum consists of the two-boson continuum, whose eigenstates $|k_1,k_2\rangle$ are 
given by (\ref{eq:basis1}). The eigenenergies are the sum of the two single-particle energies:
\begin{equation}\label{eq:energyunperturbed}
E_{k_1,k_2}^0=-2[\cos(k_1)+\cos(k_2)].
\end{equation}
When $\gamma>0$, eigenvalues in the lower part of the spectrum are pushed down, and beyond $\gamma\approx2$ a band of $f$ states splits off from the two-boson continuum. These are the two-boson bound states, with a high probability of finding the two bosons on the same lattice site, while the probability
of them being separated by a distance $r$ decreases exponentially with increasing $r$
\cite{EilbeckPhysicaD78,Scott1,NguenangPRB75}.

The critical value $\gamma_b=2$ for which the band of two-boson bound states splits off from the 
continuum may be explained as follows. In the limit $f\to\infty$ the unnormalized bound state with highest energy $E=-2\gamma$ is given by \cite{NguenangPRB75,Eilbeck03}:
\begin{equation}
|\Psi\rangle = \sum_{l=1}^f (-1)^l |l,l\rangle.
\end{equation}
For $\gamma_b=2$ the energy of that state leaves the two-boson continuum of energies $E\in [-4,4]$.

\section{Localization in normal-mode space}\label{localization}

We recall that the normal-mode space is spanned by both momenta $k_1$ and 
$k_2$. The conditions $0<k_{1,2}<\pi$ and $k_1\leq k_2$ reduce 
the normal-mode space to a triangle that we call {\it the irreducible triangle}, as sketched in 
Fig. \ref{clines}. For finite $f$ and $\gamma$ the eigenstates $|\Psi\rangle$ will spread in the basis of the $\gamma=0$
eigenstates $\{|k_1,k_2\rangle\}$. We measure such a spreading by computing the weight function in 
normal-mode space $C(k_1,k_2)=|\langle k_1,k_2|\Psi\rangle|^2$.

\subsection{Analysis by perturbation theory}

\begin{figure}[!t]
\begin{center}
\includegraphics[width=2.2in]{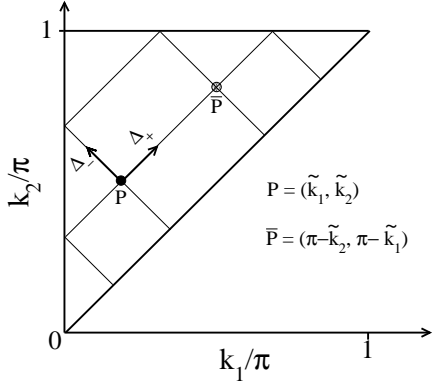}
\end{center}
\caption{\label{clines}Sketch of the different lines in the two-dimensional normal-mode 
space along which the weight function (\ref{eq:weight}) is nonzero. The seed point $P=(\tilde{k}_1,\tilde{k}_2)$ corresponding to the unperturbed eigenstate $|\tilde{k}_1,\tilde{k}_2\rangle$ is 
represented by the black spot. Its conjugate point $\bar{P}=(\pi-\tilde{k}_2,\pi-\tilde{k}_1)$ 
is represented by the grey spot. The axes defining the coordinates $\Delta_+$ and $\Delta_-$ 
are indicated by the arrows emerging from $P$.}
\end{figure}

We use perturbation theory to calculate the weight functions, where $\gamma$ is 
the perturbation. We fix the momentum $k_1$ and $k_2$, and choose an 
eigenstate $|\tilde{k}_1,\tilde{k}_2\rangle$ of the unperturbed case $\gamma=0$. The wave numbers $\tilde{k}_1$ and $\tilde{k}_2$ define a {\it seed point} $P = (\tilde{k}_1,\tilde{k}_2)$ in the irreducible triangle (see  Fig. \ref{clines}).
Upon 
increase of $\gamma$, the chosen eigenstate transforms into a new eigenstate $|\Psi_{\tilde{k}_1\tilde{k}_2}\rangle$,
which will have overlap with several eigenstates of the $\gamma=0$ case. We 
expand the eigenfunction of the perturbed system to first order in $\gamma$:
\begin{equation}
|\Psi_{\tilde{k}_1\tilde{k}_2}\rangle = |\tilde{k}_1,\tilde{k}_2\rangle 
+ \gamma |\Psi_{\tilde{k}_1,\tilde{k}_2}^{(1)}\rangle,
\end{equation}
where
\begin{equation}
|\Psi_{\tilde{k}_1,\tilde{k}_2}^{(1)}\rangle = \sum_{k'_1\neq\tilde{k}_1}\sum_{\substack{k'_2\neq\tilde{k}_2 \\ k'_2\ge k'_1}} \frac{\langle k'_1,k'_2|\hat{H}_1|\tilde{k}_1,\tilde{k}_2\rangle}{{E_{{\tilde{k}_1}{\tilde{k}_2}}^0}-E_{{{k}_1^{\prime}}{k_2^{\prime}}}^0} |k'_1,k'_2\rangle .
\end{equation}
Thus for $k_1\neq\tilde{k}_1$ and $k_2\neq\tilde{k}_2$ the weight function 
$C(k_1,{k}_2;{\tilde{k}_1},\tilde{k}_2) =  |\langle k_1,k_2|\Psi_{\tilde{k}_1{\tilde{k}_2}}\rangle|^2$ is
\begin{equation}\label{eq:weight}
C(k_1,{k}_2;{\tilde{k}_1},\tilde{k}_2) = \gamma^2
\frac{|\langle k_1,k_2|\hat{H}_1|\tilde{k}_1,\tilde{k}_2\rangle|^2}{|E_{{\tilde{k}_1}{\tilde{k}_2}}^0-E_{{k}_1{k}_2}^0|^2},
\end{equation}
where $E_{k_1k_2}^0$ and $E_{\tilde{k}_1\tilde{k}_2}^0$ are eigenenergies of the unperturbed
system given by (\ref{eq:energyunperturbed}). For convenience we use new variables in normal-mode space
\begin{equation}
k_{\pm} = k_2 \pm k_1, \label{eq:kpluskminus1}
\end{equation}
which are the total (Bloch) and relative wave numbers respectively. They have values 
$0< k_+< 2\pi$ and $0< k_-< \pi$.
Since we are interested in the behavior of the weight function around the core at 
$(\tilde{k}_1,\tilde{k}_2)$, we define the coordinates relative to that point: 
\begin{equation}\label{eq:deltaminus}
\Delta_{\pm} = {k}_{\pm} - \tilde{k}_{\pm}.
\end{equation}
Thus, (\ref{eq:weight}) becomes
\begin{eqnarray}\label{eq:weight2}
C(k_1,k_2;\tilde{k}_1,\tilde{k}_2) &=& \gamma^2  \frac{f_{q_1,q_2,\tilde{q}_1,\tilde{q}_2}^2}
{[16(E_{\tilde{k}_1\tilde{k}_2}^0-E_{k_1 k_2}^0)]^2} \nonumber \\
 & & \;\;\;\;\;\;\;\;\;\;\;\;\; \times \; R_{k_+,k_-;\tilde{k}_+,\tilde{k}_-}^2,
\end{eqnarray}
where $f_{q_1,q_2,\tilde{q}_1,\tilde{q}_2}$ is given by Eq. (\ref{eq:fq}).

The coefficient $R_{k_+,k_-;\tilde{k}_+,\tilde{k}_-}$ consists of a sum of eight terms of the form
\begin{equation}\label{eq:gfunction}
g(\zeta) = \frac{\sin\left[(2f+1)\frac{\zeta}{2}\right]}{\sin\left(\frac{\zeta}{2}\right)},
\end{equation}
with pairwise opposite signs (see appendix \ref{applines}). 
For each term, the argument $\zeta$ is a certain combination of the wave numbers $k_+,k_-$ and 
$\tilde{k}_+,\tilde{k}_-$ (see appendix \ref{applines} for details). Unless the argument of any of the eight 
terms $g(\zeta)$ vanishes, all of them cancel each other and $R_{k_+,k_-;\tilde{k}_+,\tilde{k}_-}=0$.
Thus the condition $\zeta=0$ for each term in $R_{k_+,k_-;\tilde{k}_+,\tilde{k}_-}$, together with 
the relations (\ref{eq:kpluskminus1}) and (\ref{eq:deltaminus}), defines lines $k_2=k_2(k_1)$ in 
the normal-mode space where the weight function $C(k_1,k_2;\tilde{k}_1,\tilde{k}_2)$ is nonzero. 
These lines are schematically shown in Fig. \ref{clines} (the analytical derivation of these lines 
is given in appendix \ref{applines}). Note that these lines are specularly reflected at the 
boundaries $k_1=0$ and $k_2=\pi$ of the irreducible triangle.

To study the localization in normal-mode space away from the core using the formula (\ref{eq:weight2}), 
we consider the two cases $\Delta_{-}=0$, $\Delta_{+}>0$ and vice versa, i.e. the mutually perpendicular 
directions $\Delta_+$ and $\Delta_-$ (Fig. \ref{clines}). For each case we obtain,
with $|\Delta_{\pm}|\ll \pi$,
\begin{eqnarray}\label{eq:weightformula}
C^{\pm}(k_1,{k}_2;{\tilde{k}_1},\tilde{k}_2) &=& \left(\frac{\gamma}{f+1}\right)^2 (2-{\delta}_{q_1,q_2})
(2-{\delta}_{\tilde{q}_1,\tilde{q}_2}) \nonumber \\ &\times & \Delta_{\pm}^{-2}
\Bigg\{\left[\cos(\tilde{k}_1)+\cos(\tilde{k}_2)\right]\frac{\Delta_{\pm}}{2} \nonumber \\ &+&\sin(\tilde{k}_1)\pm \sin(\tilde{k}_2)\Bigg\}^{-2}.
\end{eqnarray}
The effective interaction strength is $\gamma/(f+1)$. In the limit $\gamma\to 0$ or 
$f\to \infty$ we have compactification of the eigenstates.
The formula (\ref{eq:weightformula}) shows localization in normal-mode space. Depending on the seed $(\tilde{k}_1,\tilde{k}_2)$ we find algebraic decay within the irreducible triangle, $C\sim \Delta^{-\alpha}$, with $\alpha = 2,4$. If $\sin\tilde{k}_1 \pm \sin\tilde{k}_2 \neq 0$, $\alpha=2$. If $\sin\tilde{k}_1 \pm \sin\tilde{k}_2 = 0$, $\alpha=4$. E.g. for $\tilde{k}_1 = \tilde{k}_2$
\begin{equation}\label{eq:delta42}
C^{-}\sim \left(\frac{\gamma}{f+1}\right)^2\frac{1}{\cos^2(\tilde{k}_1)\Delta_{-}^4}.
\end{equation}

Note that along the $\Delta_{+}$ direction in the irreducible triangle, 
$R_{k_+,k_-;\tilde{k}_+,\tilde{k}_-}=2(f+1)$ at all points but $\bar{P}=(\bar{k}_1=\pi-\tilde{k}_2,\bar{k}_2=\pi-\tilde{k}_1)$. 
This is the conjugate point of the seed $P$ (Fig. \ref{clines}), where two lines intersect. At this point 
$R_{k_+,k_-;\tilde{k}_+,\tilde{k}_-}=4(f+1)$. Thus we expect a local maximum of the weight function 
at the conjugate point. The states $|\tilde{k}_1,\tilde{k}_2\rangle$ and $|\bar{k}_1,\bar{k}_2\rangle$ have energies $E_{\bar{k}_1,\bar{k}_2}^0 = -E_{\tilde{k}_1,\tilde{k}_2}^0$.

\subsection{Numerical results}

\begin{figure}[!t]
\begin{center}
\includegraphics[width=2.3in]{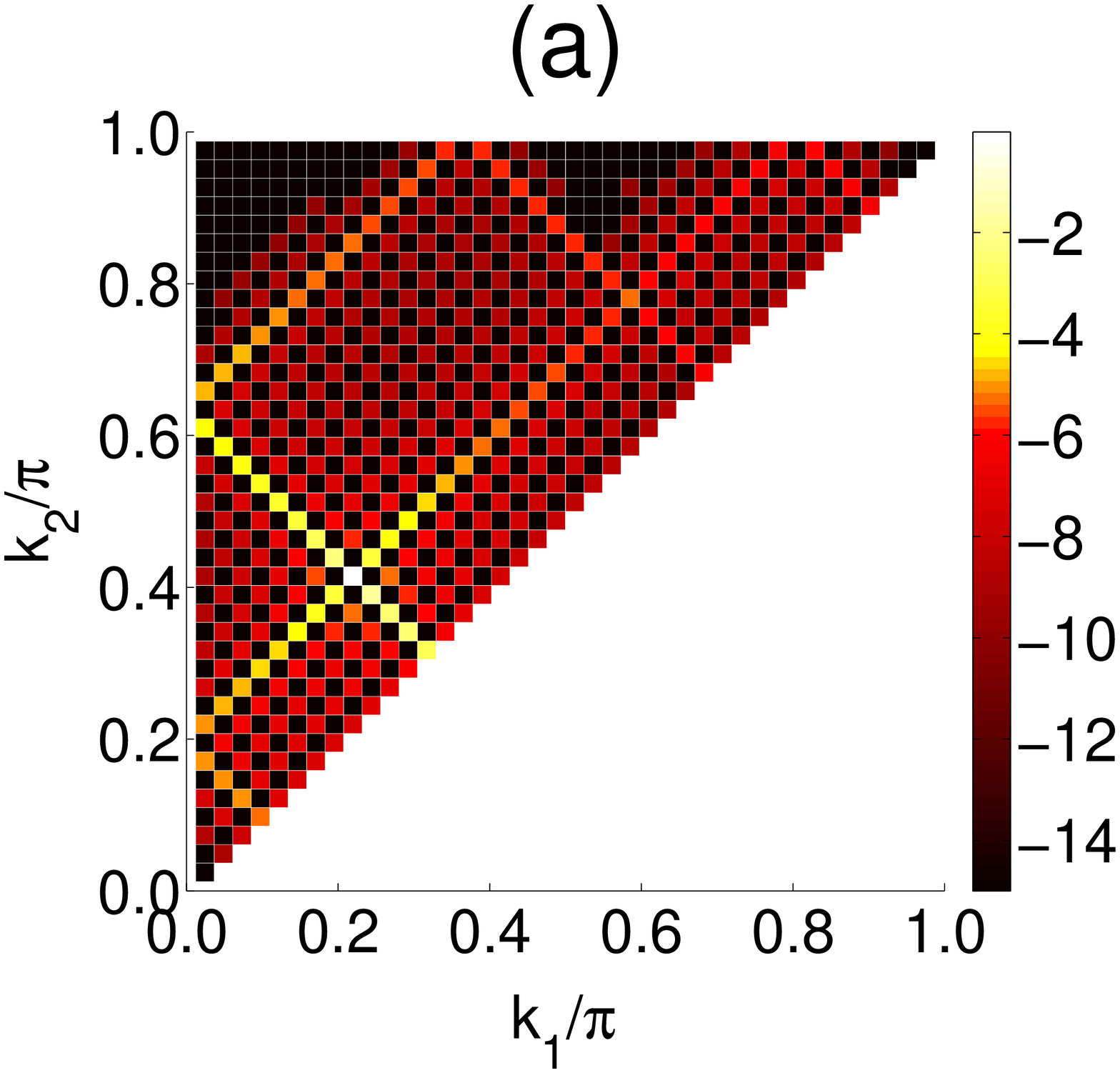}
\includegraphics[width=2.3in]{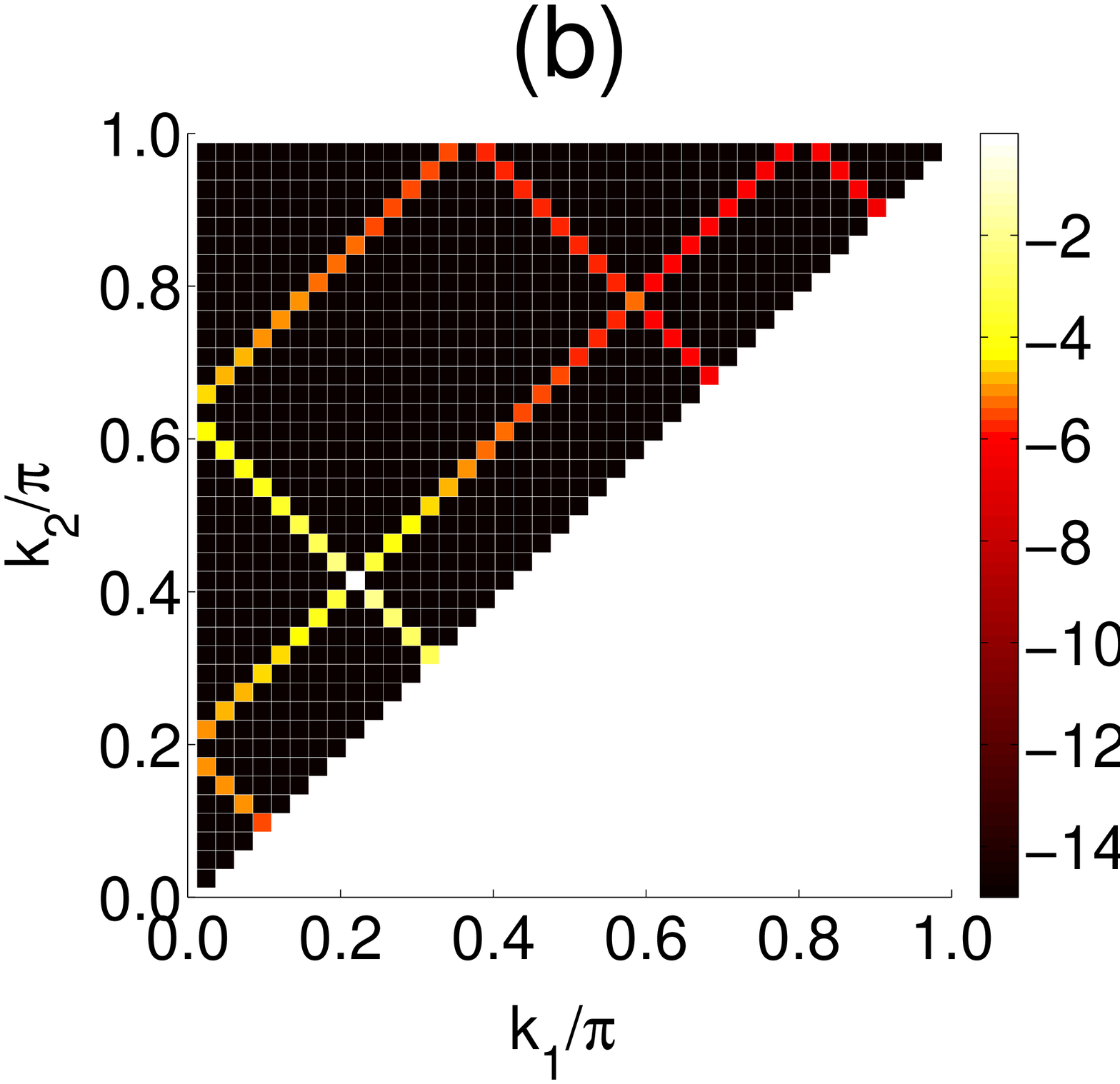}
\end{center}
\caption{\label{scatter}3-D plot of the logarithm of the weight function in the normal-mode 
space for the eigenstate $\nu=145$, obtained by (a) numerical diagonalization, and (b) perturbation 
theory using the formula (\ref{eq:weight2}). Here $f=40$ and $\gamma=0.1$.}
\end{figure} 
In Fig. \ref{scatter} we show the weight function in the two-dimensional normal-mode space 
obtained by numerical diagonalization and the formula (\ref{eq:weight2}) respectively, with
characteristic localization profiles. We find agreement of the numerical data 
with the results 
from perturbation theory. The largest value is at the point 
$P=(\frac{9}{40}\pi,\frac{17}{40}\pi)\sim(0.2\pi,0.4\pi)$, and it decays mainly along the lines described 
in the previous section (Fig. \ref{clines}). Note also the presence of the local maximum at the 
conjugate point $\bar{P}\sim(0.6\pi,0.8\pi)$ in both cases.
\begin{figure}[!t]
\begin{center}
\includegraphics[width=2.6in]{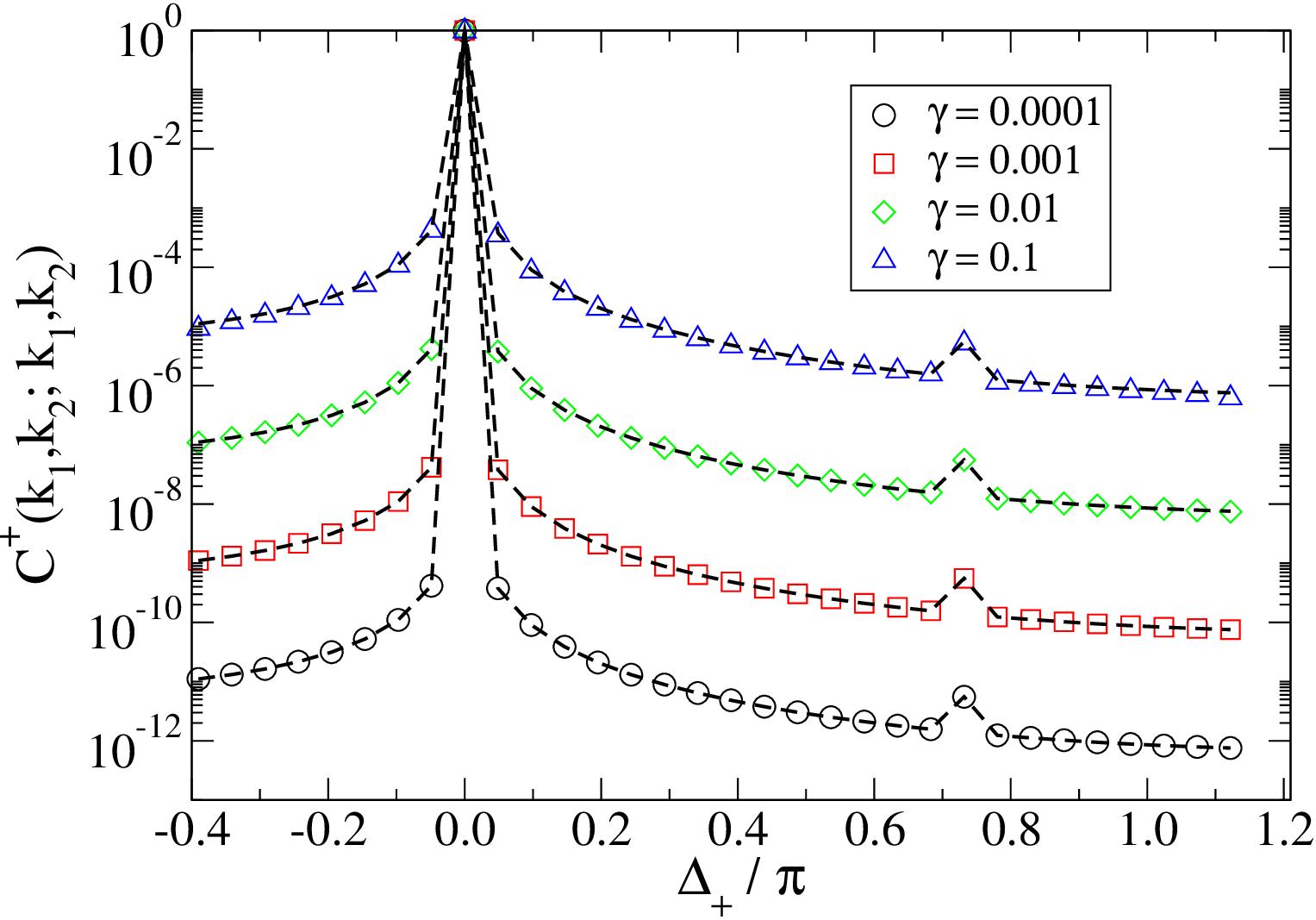}
\end{center}
\caption{\label{dplus}Weight function for different values of the interaction strength $\gamma$ of 
the eigenstate $\nu=145$ along the $\Delta_+$ direction. The dashed lines are results from formula 
(\ref{eq:weight2}). Here $f=40$.}
\end{figure} 
\begin{figure}[!t]
\begin{center}
\includegraphics[width=2.6in]{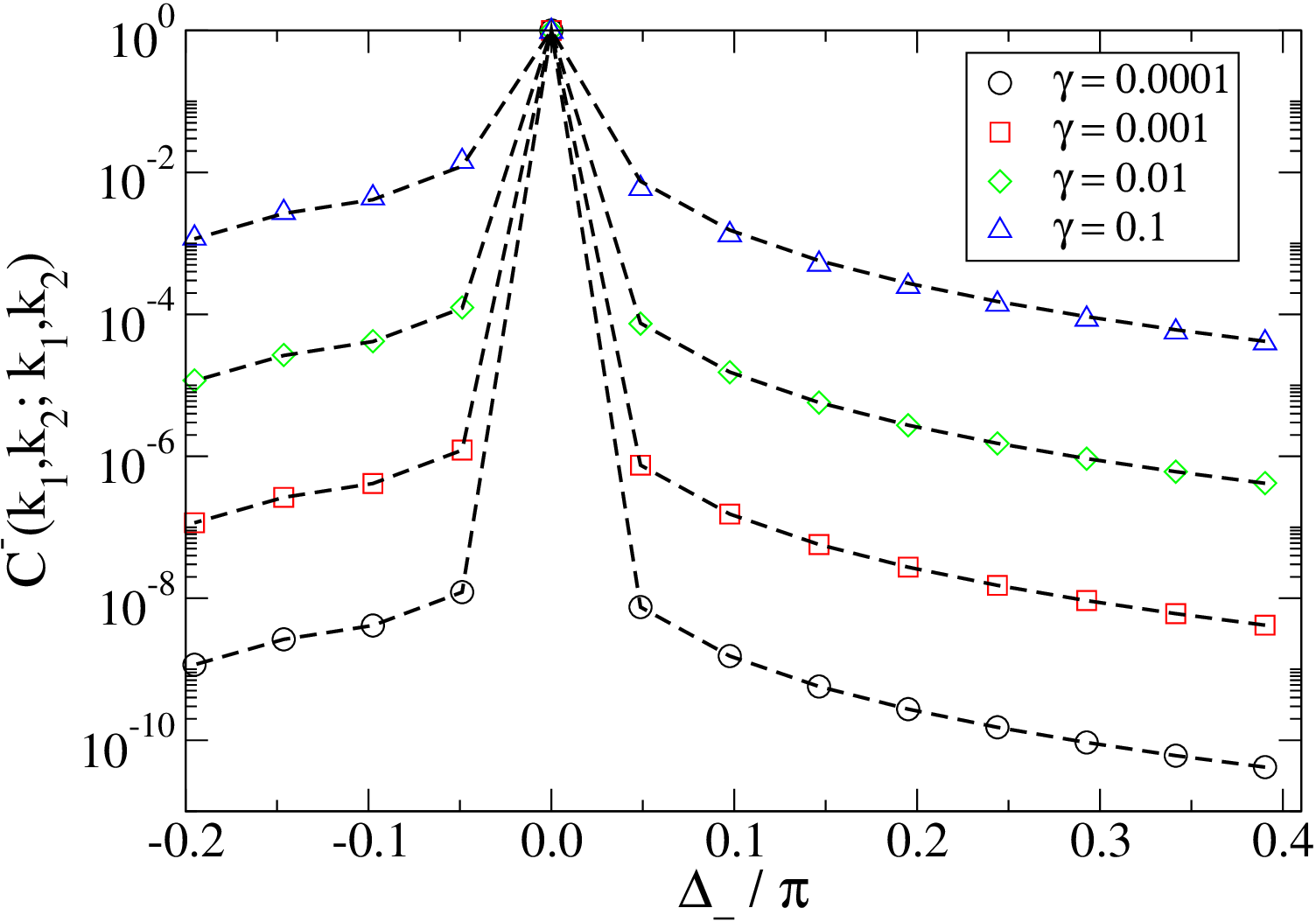}
\end{center}
\caption{\label{dminus}Weight function for different values of the interaction strength $\gamma$ of 
the eigenstate $\nu=145$ along the $\Delta_-$ direction. The dashed lines are results from formula 
(\ref{eq:weight2}). Here $f=40$.}
\end{figure} 

In Figs. \ref{dplus} and \ref{dminus} we plot the weight function of the eigenstate shown in 
Fig. \ref{scatter} along the directions $\Delta_+$ and $\Delta_-$ respectively for different values 
of the interaction parameter $\gamma$. The state becomes less localized with increasing $\gamma$, 
as expected from the above analysis. The decay of the weight function is well 
described by perturbation theory (dashed lines). The peak of the weight function at the conjugate point is clearly seen in Fig. \ref{dplus}.

\begin{figure}[!t]
\begin{center}
\includegraphics[width=2.6in]{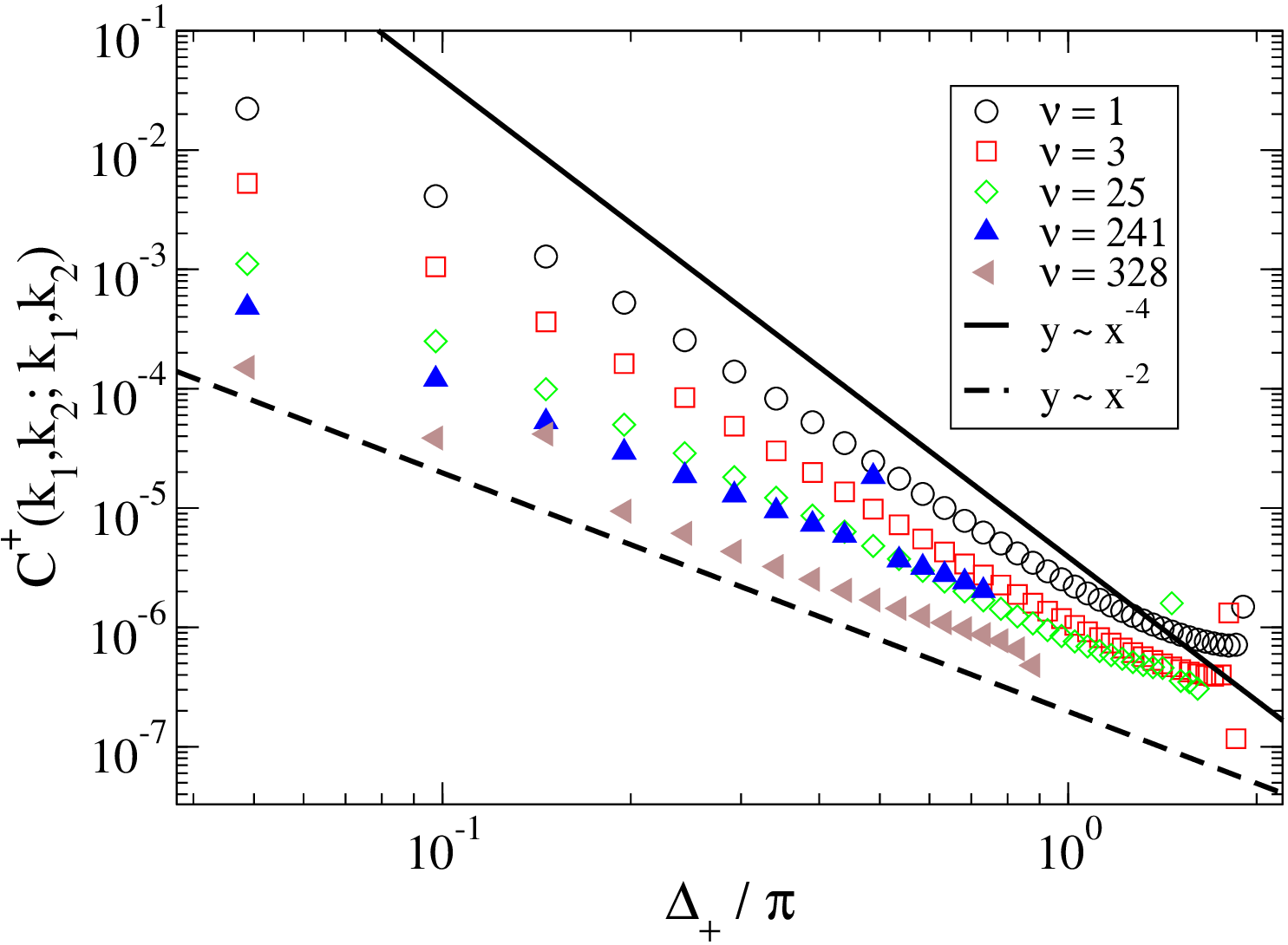}
\end{center}
\caption{\label{dpluslog}Weight function of different eigenstates (labeled by the index $\nu$) along the $\Delta_+$ direction. 
Here $\gamma=0.1$ and $f=40$.}
\end{figure}

In Fig. \ref{dpluslog} we plot the weight function of different states
along the $\Delta_+$ 
direction. It decays as a power law that ranges from $\Delta^{-4}$ for states near the 
lower corner of the irreducible triangle (see Fig. \ref{kpoints}) to $\Delta^{-2}$ for states fulfilling 
$\tilde{k}_2\approx \pi - \tilde{k}_1$. In Fig. \ref{dminuslog} we plot the decay of the weight function 
along the $\Delta_-$ direction, where we see the power-law decay that ranges from $\Delta^{-4}$ for states fulfilling $\tilde{k}_1\approx\tilde{k}_2$ (see Fig. \ref{kpoints}), to $\Delta^{-2}$ for 
states fulfilling $\tilde{k}_2\approx \pi - \tilde{k}_1$. The results from numerical diagonalization agree very well with those from the perturbation theory analysis.
\begin{figure}[!t]
\begin{center}
\includegraphics[width=2.6in]{Figure7.eps}
\end{center}
\caption{\label{dminuslog}Weight function of different eigenstates (labeled by the index $\nu$) along the $\Delta_-$ direction. 
Here $\gamma=0.1$ and $f=40$.}
\end{figure} 
\begin{figure}[!t]
\begin{center}
\includegraphics[width=2.in]{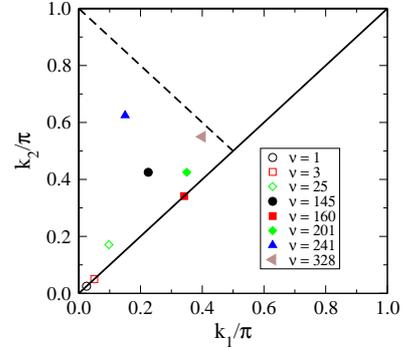}
\end{center}
\caption{\label{kpoints}Location $P=(\hat{k}_1,\hat{k}_2)$ of the eigenstates, shown in 
Figs. \ref{dpluslog} and \ref{dminuslog}, in the irreducible triangle.  
}
\end{figure} 

\subsection{Nonperturbative effects}

\begin{figure}[!t]
\begin{center}
\includegraphics[width=2.3in]{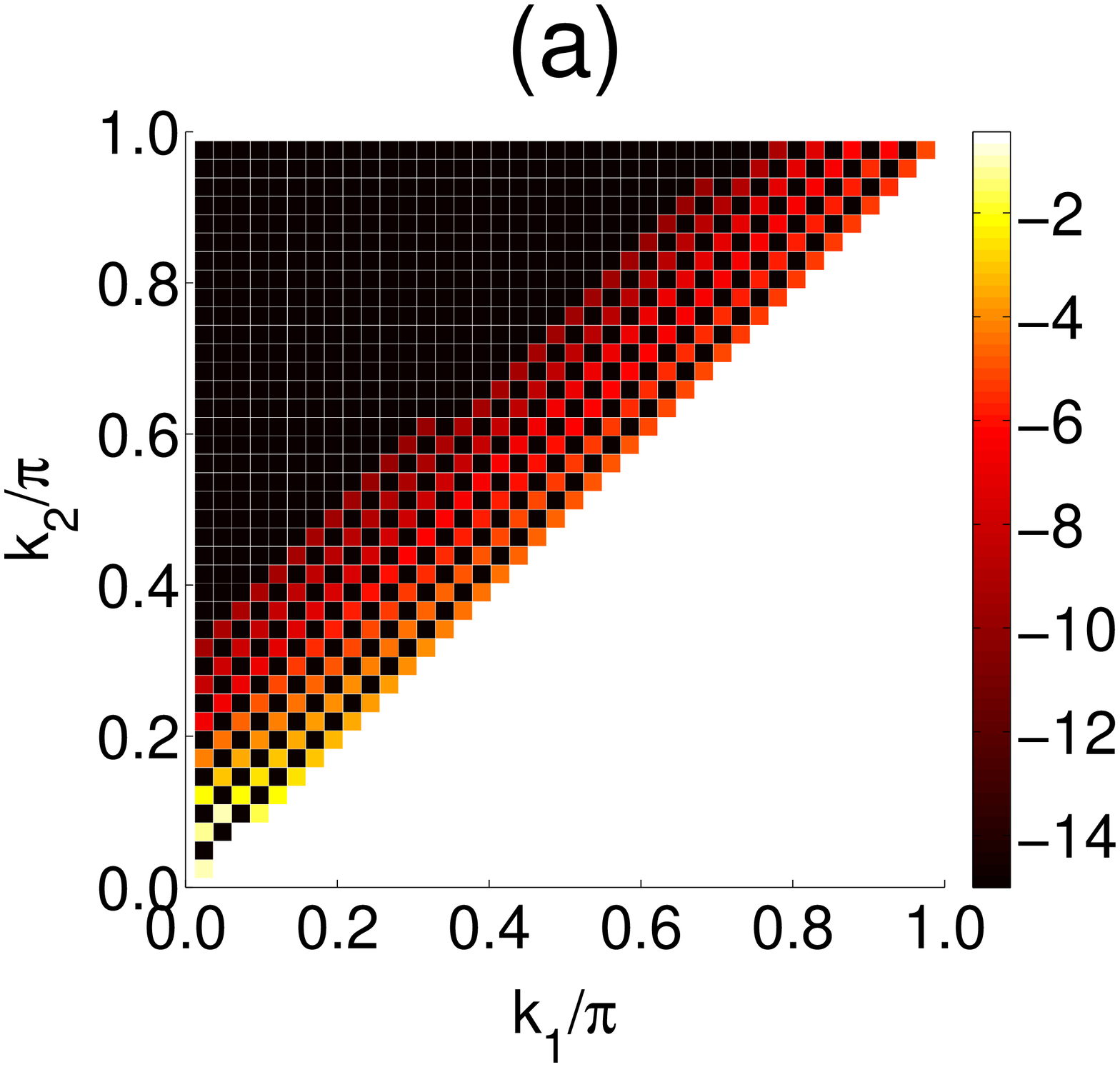}
\includegraphics[width=2.3in]{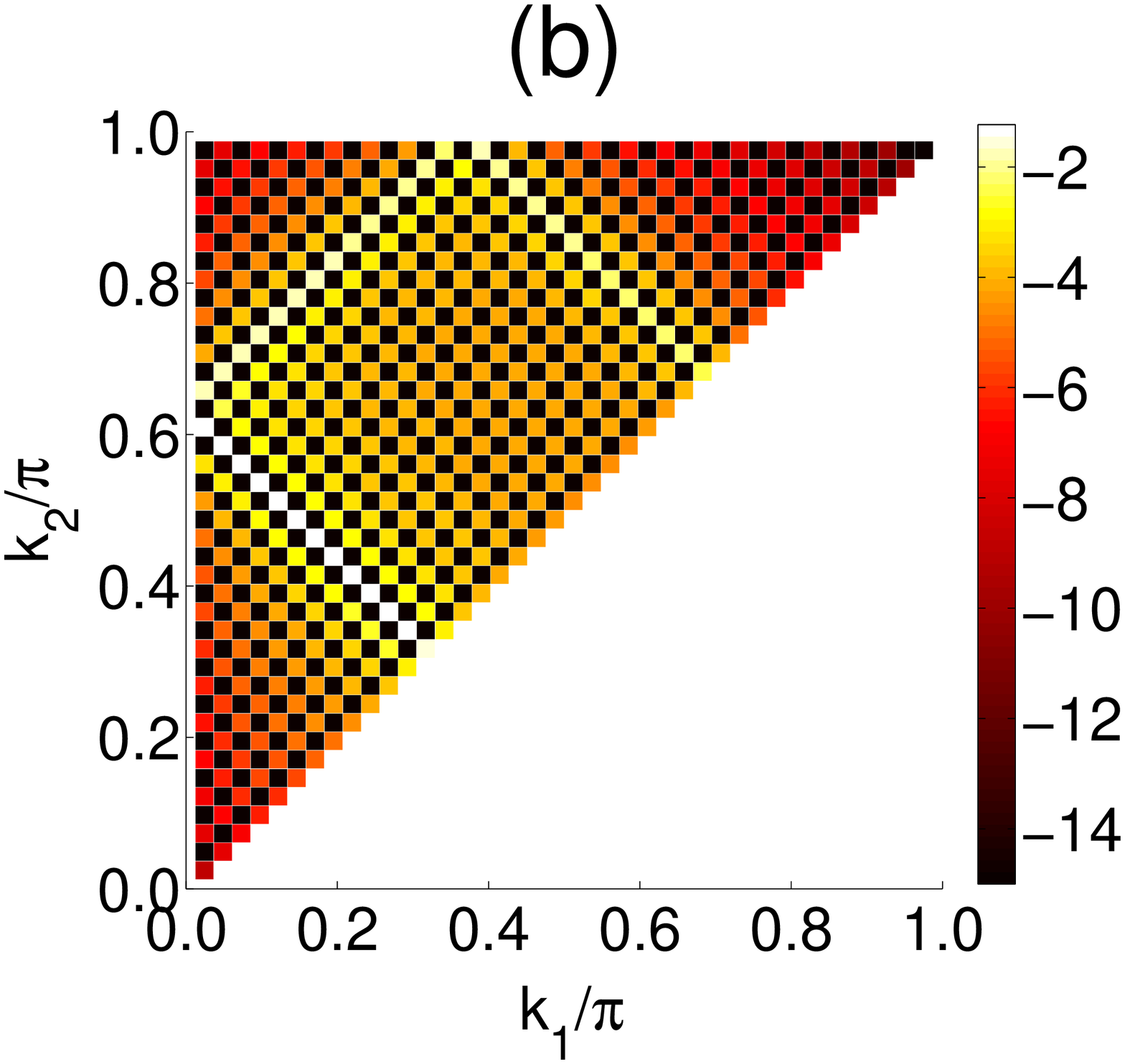}
\end{center}
\caption{\label{delocalized}3-D plot of the logarithm of the weight function in normal-mode space for the eigenstate 
$\nu=25$, that belongs to the two-boson bound state band. Results were obtained by numerical diagonalization. (a) $\gamma=1.5$. 
The two-boson bound state band did not split off and the eigenstate is localized in normal-mode space. (b) $\gamma=2$. 
At this interaction value the two-boson bound state band splits off and the eigenstate becomes 
delocalized in normal-mode space. Here $f=40$.}
\end{figure}

The results in the previous section were obtained for small values of the interaction parameter $\gamma$ up to $\gamma=0.1$, 
for which perturbation theory gives a good description of the results obtained by numerical 
diagonalization. However, when increasing $\gamma$ several nonperturbative effects occur. These are:

\underline{{\it Split off of the two-boson bound state band}}: This effect was discussed in Sec. \ref{twobosonstates} 
(Figs. \ref{spectrum}). When $\gamma >2$ the two-boson bound state band splits off from the two-boson 
continuum, and the corresponding eigenstates are correlated in direct space, i.e. with large probability
the two bosons are occupying identical lattice sites. Thus, in 
normal-mode space these eigenstates become delocalized as shown in Fig. \ref{delocalized}. 

\begin{figure}[!t]
\begin{center}
\includegraphics[width=1.55in]{Figure10a.eps}
\includegraphics[width=2.3in]{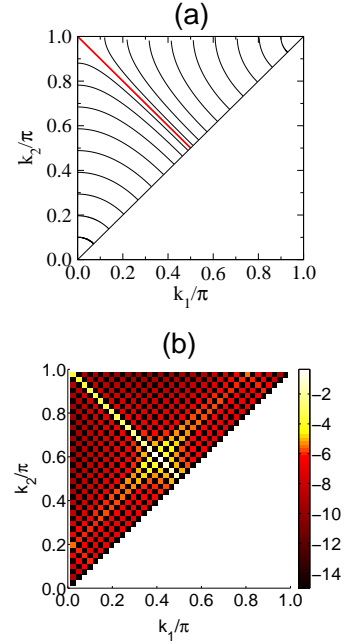}
\end{center}
\caption{\label{lines}(a) Lines of constant energy given by (\ref{eq:energyunperturbed}). The red thick line is the 
$E_{k_1,k_2}^{0} = 0$-line where all unperturbed states $|k_1,k_2 = \pi - k_1\rangle$ are degenerate. 
(b) 3-D plot of the logarithm of the weight function in normal-mode space for the eigenstate $\nu=405$, 
obtained by numerical diagonalization with $\gamma=0.1$ and $f=40$. In the noninteracting case this state corresponds to
$|k_1,k_2 = \pi - k_1\rangle$.}
\end{figure}

\underline{{\it Degenerate levels in the noninteracting case}}: The analysis using perturbation theory is valid as long as the
eigenstate which is continued from the noninteracting case is not degenerate. Because of the finiteness of the lattice the momenta $\tilde{k}_1$ and $\tilde{k}_2$ are restricted to discrete values and define a grid in the two-dimensional normal-mode space. 
A grid point $(\tilde{k}_1,\tilde{k}_2)$ defines a line of constant energy in normal-mode space through 
Eq. (\ref{eq:energyunperturbed}), with $E_{k_1,k_2}^0 = E_{\tilde{k}_1,\tilde{k}_2}^0$ (Fig. \ref{lines}-a). 
The nondegeneracy condition implies that this line should not pass through any other grid point. It is easy to see from Eq. (\ref{eq:energyunperturbed}) that all states $|\tilde{k}_1,\pi-\tilde{k}_1\rangle$ are degenerate, with $E_{\tilde{k}_1,\pi-\tilde{k}_1}=0$. Their corresponding grid points in the irreducible triangle lie on the diagonal $k_2=\pi-k_1$ (thick line in Fig. \ref{lines}-a). In Fig. \ref{lines}-b we show the weight function of an eigenstate that is located on that diagonal in the noninteracting case. As expected, even for small values of $\gamma$, the state completely delocalizes along the degeneracy diagonal.

\underline{{\it Avoided crossings}}: Upon increase of the interaction parameter $\gamma$, 
the energies of continued eigenstates change, and will resonate with eigenvalues of other states.

The first possible avoided level crossing defines a critical value of the interaction parameter $\gamma$ up to which first-order 
perturbation theory is applicable. 
To estimate this value, $\gamma_c$, we assume that before the first avoided crossing is encountered, 
the eigenenergies depend linearly on 
$\gamma$. This dependence may be estimated using first-order perturbation theory in $\gamma$. The result is, for large $f$,
\begin{equation}\label{eq:Evsgamma}
E_{\tilde{k}_1,\tilde{k}_2}(\gamma) \approx E_{\tilde{k}_1,\tilde{k}_2}^0 + \frac{b(\tilde{k}_1,\tilde{k}_2)}{f} \gamma,
\end{equation}
where
\begin{equation}\label{eq:b}
b(\tilde{k}_1,\tilde{k}_2) = \left\{ \begin{array}{ll}
2 & \textrm{if $\tilde{k}_1=0$},\\
-2 & \textrm{if $\tilde{k}_2=\tilde{k}_1>0$},\\
-1 & \textrm{if $\tilde{k}_1>0,\; \tilde{k}_2>\tilde{k}_1$.}
\end{array} \right.
\end{equation}

Let us consider two levels $E_1$ and $E_2$ that interact in the first avoided level crossing. 
At $\gamma=0$ they are separated by $\delta E$. For nonzero $\gamma$ the energies linearly change in $\gamma$:
\begin{equation}
E_1 = E_{\tilde{k}_1,\tilde{k}_2}^0 + \frac{|b_1(\tilde{k}_1,\tilde{k}_2)|}{f} \gamma,
\end{equation}
\begin{equation}
E_2 = E_{\tilde{k}_1,\tilde{k}_2}^0 + \delta E - \frac{|b_2(\tilde{k}_1,\tilde{k}_2)|}{f} \gamma.
\end{equation}
By equating $E_1$ and $E_2$ at $\gamma=\gamma_c$ we obtain
\begin{equation}
\gamma_c(\tilde{k}_1,\tilde{k}_2) = \frac{f\delta E}{|b_1(\tilde{k}_1,\tilde{k}_2)| + |b_2(\tilde{k}_1,\tilde{k}_2)|}.
\end{equation}

The first avoided crossing of the level $E_1$ will happen with its nearest neighbor in the spectrum of the $\gamma=0$ case, 
which is separated by $\delta E(\tilde{k}_1,\tilde{k}_2)$. Using 
Eq. (\ref{eq:energyunperturbed}) with $\tilde{k}_1,\tilde{k}_2 \ll \pi/2$ (see Fig. \ref{lines}-a), the separation 
is estimated (see appendix \ref{appde}): $\delta E \approx 4\sqrt{2}\pi/(f+1)^2$. Therefore $\gamma_c\sim 1/f$.

The coefficient $b_1(\tilde{k}_1,\tilde{k}_2)$ depends on the state $|\tilde{k}_1,\tilde{k}_2\rangle$ under consideration 
through (\ref{eq:b}). The coefficient $b_2(\tilde{k}_1,\tilde{k}_2)$ must have opposite sign as compared to $b_1(\tilde{k}_1,\tilde{k}_2)$ 
for the avoided crossing to take place. For the states $\nu=145$ and $\nu=41$ located at 
$(\tilde{k}_1,\tilde{k}_2)\approx (0.2\pi,0.4\pi)$ and $(\tilde{k}_1,\tilde{k}_2)\approx (0.12\pi,0.22\pi)$ respectively 
(see Fig. \ref{kpoints}), $b_1(\tilde{k}_1,\tilde{k}_2) = -1$ and $b_2(\tilde{k}_1,\tilde{k}_2) = 2$. 
This leads to a critical value of the interaction parameter $\gamma_c\approx 0.28$, which is in reasonable agreement with the 
numerical results: $\gamma_c\approx 0.2$ for the state $\nu=145$, and $\gamma_c\approx 0.3$ for the state $\nu=41$. 

\section{Conclusions}\label{conclusions}

In this work we studied the properties of quantum q-breathers in a
one-dimensional lattice containing two bosons modeled by the BH
Hamiltonian with fixed boundary conditions. Because of the lack of translational invariance, the normal-mode space is 
two-dimensional and reduces to a triangle when working in the irreducible representation of the product basis states 
(the irreducible triangle). To explore localization phenomena in this system we computed appropriate weight functions 
of the eigenstates in the normal-mode
space using both perturbation theory and numerical diagonalization. We find that the weight function is sizable only 
along the mutually perpendicular directions defined by the total and relative momentum, thus it defines lines in the 
irreducible triangle that show specular reflections at the boundaries of the irreducible triangle. We observe
localization of the weight function along these lines. The localization is
stronger when the size of the system increases or the interaction parameter is weaker, the former because the effective interaction drops in the dilute limit of large chains.
We found algebraic localization. The power of the decay is different for each eigenstate depending on which seed wave numbers have in the noninteracting case, ranging from two to four.

An interesting effect is the local maximum of the weight function at the symmetry-related (conjugate) point of the eigenstate core in normal-mode space, due to a crossing between different paths described by the lines along which the weight function is nonzero within perturbation theory.

In addition to the existence of degeneracies between eigenstates in the noninteracting case, we analyzed other nonperturbative effects as the interaction parameter increases, which limit the applicability of perturbation theory to describe the system: The splitting off of the two-boson bound states from the two-boson continuum, and the occurrence of avoided level crossings. The first effect manifests as a delocalization of the weight function of the bound states due to the two-boson correlation in direct space. The second effect manifests as a sudden change of the location of an eigenstate in the normal-mode space due to resonant interaction with another eigenstate. Both effects define critical values of the interaction parameter below which one may analyze the system by perturbation theory. The occurrence of an avoided level crossing gives the smallest critical value.
 
Although we considered a system with fixed boundary conditions, we still obtain algebraic decay as
in the case with periodic boundary conditions \cite{NguenangPRB75}. The question how to restore exponential localization of 
classical q-breathers from algebraic decay of quantum q-breathers in the limit of large numbers of particles is still open. When going to that limit, one may use a Hartree approximation and describe the system with a product state wavefunction, or use a coherent state representation. Both ways lead to the nonlinear Schr\"odinger equation where classical q-breathers are known to exist \cite{kgm08}.

\section*{Acknowledgements}

J.P.N. acknowledges the warm hospitality of the Max
Planck Institute for the Physics of Complex Systems in Dresden. This work was
supported by the DFG (grant no. FL200/8) and by the ESF network-programme AQDJJ.

\appendix

\section{Lines of nonzero weight function}\label{applines}

For fixed $\tilde{k}_1,\tilde{k}_2$, the coefficient $R(k_1,k_2;\tilde{k}_1,\tilde{k}_2)$ in Eq. (\ref{eq:weight2}) is given by
\begin{eqnarray}\label{eq:corel}
R(k_1,k_2;\tilde{k}_1,\tilde{k}_2)  &=&  
g(k_- + \tilde{k}_-)
+ g(\Delta_-)
\nonumber \\
&-& g(k_- + \tilde{k}_+)
- g(k_- - \tilde{k}_+)  
\nonumber \\
&-& g(k_+ + \tilde{k}_-)
- g(k_+ - \tilde{k}_-)
\nonumber \\
&+& g(\Delta_+)
+ g(k_+ + \tilde{k}_+),
\end{eqnarray}
where
\begin{equation}
g(\zeta) = \frac{\sin[(2f+1)\frac{\zeta}{2}]}{\sin(\frac{\zeta}{2})}.
\end{equation}
The lines $k_2=k_2(k_1)$ in the normal-mode space (irreducible triangle) along which $R(k_1,k_2;\tilde{k}_1,\tilde{k}_2)\neq 0$ 
are obtained from the condition that the argument of any term in (\ref{eq:corel}) is zero, such that
\begin{equation}
g(\zeta) = 2f + 1.
\end{equation}
Let us analyze each of the arguments in Eq. (\ref{eq:corel}):

\begin{itemize}

\item $k_- + \tilde{k}_- = 0$: This implies that
\begin{equation}
k_2-k_1=-(\tilde{k}_2-\tilde{k}_1).
\end{equation}
Since $k_2\geq k_1$ the above condition is possible only for points $(\tilde{k}_1,\tilde{k}_2),(k_1,k_2)$ on the diagonal $k_2=k_1$. 

\item $\Delta_- = 0$: This condition leads to
\begin{equation}
k_2 = (\tilde{k}_2-\tilde{k}_1) + k_1,
\end{equation}
which is the equation of the line along the $\Delta_+$ direction that cuts the $k_2$ axis at $k_2(0)=\tilde{k}_2-\tilde{k}_1$.

\item $k_- + \tilde{k}_+ = 0$: This implies that
\begin{equation}
k_2  - k_1 = -(\tilde{k}_2+\tilde{k}_1),
\end{equation}
which is possible only if $\tilde{k}_1=\tilde{k}_2=0$ and $(k_1,k_2)$ is on the diagonal $k_2=k_1$.

\item $k_- - \tilde{k}_+ = 0$: This leads to the equation
\begin{equation}
k_2 = (\tilde{k}_2+\tilde{k}_1) + k_1,
\end{equation}
which describes a line parallel to the $\Delta_+$ direction that cuts the $k_2$ axis at $k_2(0)=\tilde{k}_2+\tilde{k}_1$.

\item $k_+ + \tilde{k}_- =0$: This implies that
\begin{equation}
k_2 = -(\tilde{k}_2+\tilde{k}_1) - k_1,
\end{equation}
which is valid only if $k_1=k_2=\tilde{k}_1=\tilde{k}_2=0$.

\item $k_+ - \tilde{k}_-=0$: This leads to the equation
\begin{equation}
k_2 = (\tilde{k}_2-\tilde{k}_1) - k_1,
\end{equation}
which is the equation of a line parallel to the $\Delta_-$ direction that cuts the $k_2$ axis at $k_2(0)=\tilde{k}_2-\tilde{k}_1$.

\item $\Delta_+=0$: This leads to the equation
\begin{equation}
k_2 = (\tilde{k}_2+\tilde{k}_1) - k_1,
\end{equation}
which describes the line along to the $\Delta_-$ direction that cuts the $k_2$ axis at $k_2(0)=\tilde{k}_2+\tilde{k}_1$.

\end{itemize}

\section{Energy separation between nearest-neighbor levels in the noninteracting case}\label{appde}

\begin{figure}[!t]
\begin{center}
\includegraphics[width=2.7in]{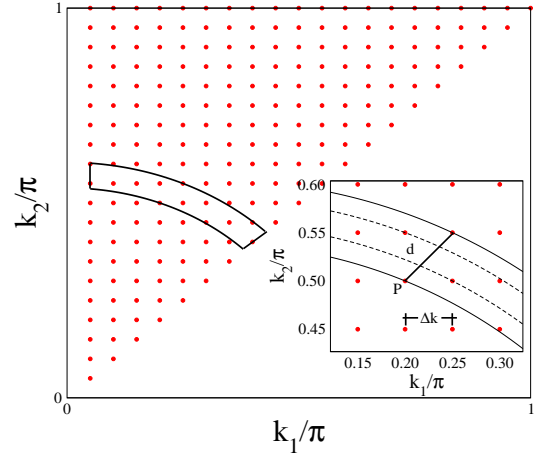}
\end{center}
\caption{\label{grid}Sketch of the discrete normal-mode space (irreducible triangle) with a line of constant energy in the circular approximation 
passing through the grid point $P=(\tilde{k}_1,\tilde{k}_2)$ and the grid point $(\tilde{k}_1+\Delta k,\tilde{k}_2+\Delta k)$. The strip of width $d=\sqrt{2}\Delta k$ and area $A_s$ 
contains $N_g=A_s/\Delta k^2$ grid points through which lines of constant energy pass. The typical line separation within the strip is $\delta k\approx d/N_g$. $\Delta k=\pi/(f+1)$ is the grid spacing.}
\end{figure} 

The finite size of the lattice leads to discrete values of the momenta $k_1$ and $k_2$, and thus to a grid in the normal-mode space (Fig. \ref{grid}). Let us consider a line of constant energy which passes
through the seed point $P=(\tilde{k}_1,\tilde{k}_2)$ given by 
$E_{k_1,k_2}^0 = E_{\tilde{k}_1,\tilde{k}_2}^0$ with Eq. (\ref{eq:energyunperturbed}). For small values of $k_1$ and 
$k_2$, the energy  in Eq. (\ref{eq:energyunperturbed}) may be approximated to
\begin{equation}\label{eq:ecircular}
E_{k_1,k_2}^0 \approx -4 + k_1^2 + k_2^2,
\end{equation}
which is the equation for a circle ({\it circular approximation}). So the equation for the line of constant energy 
passing through the point $P$ is
\begin{equation}\label{eq:k2line}
k_2(k_1;\tilde{k}_1,\tilde{k}_2) = \sqrt{\tilde{k}_1^2+\tilde{k}_2^2 - k_1^2}.
\end{equation}

Through another grid point at $(\tilde{k}_1+\Delta k,\tilde{k}_2+\Delta k)$, separated from $P$ by a distance $d = \sqrt{2} \Delta k \approx \Delta k$ ($\Delta k$ is the grid spacing), another line of constant energy with the form (\ref{eq:k2line}) passes (Fig. \ref{grid}), defining a strip of area $A_s$ in the irreducible triangle. The strip contains $N_g$ grid points through which lines of constant energy pass. The average line separation within the strip is $\delta k \approx d /N_g$.

The number of grid points in the strip is
\begin{equation}
N_g = \frac{A_s}{\Delta k^2}.
\end{equation}
The area of the strip is
\begin{eqnarray}
A_s &=& \frac{\pi}{8} [ (\tilde{k}_1+\Delta k)^2 + (\tilde{k}_2+\Delta k)^2 - \tilde{k}_1^2 - \tilde{k}_2^2 ] \nonumber \\
&=& \frac{\pi}{4}[ \Delta k^2 + (\tilde{k}_1 + \tilde{k}_2)\Delta k ].
\end{eqnarray}
Therefore
\begin{equation}
N_g = \frac{\pi}{4\Delta k}(\Delta k + \tilde{k}_1 + \tilde{k}_2),
\end{equation}
and hence
\begin{equation}\label{eq:dk}
\delta k(\tilde{k}_1,\tilde{k}_2) = \frac{4\Delta k^2}{\pi(\tilde{k}_1+\tilde{k}_2+\Delta k)}.
\end{equation}

The corresponding energy separation is
\begin{equation}\label{eq:de}
\delta E(\tilde{k}_1,\tilde{k}_2) = E_{\tilde{k}_1+\delta k_1,\tilde{k}_2 +\delta k_2}^0 - E_{\tilde{k}_1,\tilde{k}_2}^0,
\end{equation}
with $\delta k_1 = \delta k_2 = \delta k/\sqrt{2}$. Substituting (\ref{eq:ecircular}) and (\ref{eq:dk}) into (\ref{eq:de}) one obtains
\begin{eqnarray}\label{eq:de2}
\delta E(\tilde{k}_1,\tilde{k}_2) &=& \frac{16\Delta k^4}{\pi^2(\tilde{k}_1+\tilde{k}_2+\Delta k)^2} + \nonumber \\ \nonumber \\
& &\frac{4\sqrt{2}(\tilde{k}_1+\tilde{k}_2)\Delta k^2}{\pi(\tilde{k}_1+\tilde{k}_2+\Delta k)}.
\end{eqnarray}
For $f$ large, and $\Delta k=\pi/(f+1) \ll \tilde{k}_{1,2}$, the first term in (\ref{eq:de2}) can be neglected. Thus we are left with
\begin{equation}
\delta E(\tilde{k}_1,\tilde{k}_2) \approx \frac{4\sqrt{2}\pi}{(f+1)^2}.
\end{equation}

\end{document}